\documentstyle[12pt,axodraw,epsfig]{article}
\newcommand{\inv}{\left. g^0_A \right|_{\rm inv}}

\newcommand{\Frac}[2]%
{{\textstyle \frac{\mbox{\footnotesize $#1$}\rule[-0.9mm]{0mm}{1mm}}%
{\mbox{\footnotesize $#2$}\rule{0mm}{3.1mm}}}}
\renewcommand{\thefootnote}{\fnsymbol{footnote}}

\begin{document}
\begin{titlepage}
\vspace*{-12 mm}
\noindent
\begin{flushright}
\begin{tabular}{l@{}}
\end{tabular}
\end{flushright}
\vskip 12 mm
\begin{center}
{\Large \bf 
Gluonic effects in $\eta'$-nucleon interactions
} 
\\[10mm]
{\bf Steven D. Bass}\footnote{email: sbass@ect.it}
\\[10mm]   
{\em 
ECT*, Strada delle Tabarelle 286, I-38050 Villazzano, Trento, Italy
 }
\end{center}
\vskip 10 mm
\begin{abstract}
\noindent

We review the theory and phenomenology of the axial U(1) problem 
with emphasis on the role of gluonic degrees of freedom 
in the low-energy
$pp \rightarrow pp \eta$ and $pp \rightarrow pp \eta'$ reactions.

\end{abstract}

\vspace{2.0cm}
\begin{flushleft}
\end{flushleft} 
\end{titlepage}
\renewcommand{\labelenumi}{(\alph{enumi})}
\renewcommand{\labelenumii}{(\roman{enumii})}
\renewcommand{\thefootnote}{\arabic{footnote}}
\newpage

\section{Introduction}

$\eta$ and $\eta'$ physics together with polarised deep inelastic 
scattering provide complementary windows on the role of gluons in 
dynamical chiral symmetry breaking.
Gluonic degrees of freedom play an important role in the physics of 
the flavour-singlet $J^P = 1^+$ channel \cite{okubo} through the QCD 
axial anomaly \cite{zuoz}.
The most famous example is the $U_A(1)$ problem: the masses of the 
$\eta$ and $\eta'$ mesons are much greater than the values they 
would have if these mesons were pure Goldstone bosons associated
with spontaneously broken chiral symmetry \cite{weinberg,christos}.
This extra mass is induced by non-perturbative gluon dynamics 
\cite{thooft,hfpm,gvua1,witten} and the axial anomaly \cite{adler,bell}.

For the first time since the discovery of QCD (and the U(1) problem)
precise data are emerging on processes involving $\eta'$ production 
and decays.
There is presently a vigorous experimental programme to study the 
$pp \rightarrow pp \eta$ and $pp \rightarrow pp \eta'$ reactions
close to threshold in low-energy proton-nucleon collisions at CELSIUS 
\cite{celsius} and COSY \cite{cosy}.
New data on $\eta'$ photoproduction, $\gamma p \rightarrow p \eta'$,
are expected soon from Jefferson Laboratory \cite{cebaf} following 
earlier measurements at ELSA \cite{elsa}.
The light-mass ``exotic'' meson states with quantum numbers 
$J^{PC} = 1^{-+}$ observed at BNL \cite{exoticb} and CERN \cite{exoticc} 
in $\pi^- p$ and ${\bar p} N$
scattering were discovered in decays to $\eta \pi$ and $\eta' \pi$ 
suggesting a 
possible connection with axial U(1) dynamics.
Further ``exotic'' studies are proposed in photoproduction experiments 
at Jefferson Laboratory.
At higher energies anomalously large branching ratios have been
observed by CLEO for $B$-meson decays to an $\eta'$ plus additional 
hadrons \cite{cleob} and for the $D_s^+ \rightarrow \eta' \rho^+$ 
\cite{cleod} process.
The $B$ decay measurements have recently been confirmed in new,
more precise, data from BABAR \cite{babar} and BELLE \cite{belle}.
The LEP data on $\eta'$ production in hadronic jets is about 40\% 
short of the predictions of the string fragmentation models employed
in the JETSET and ARIADNE Monte-Carlos without an additional $\eta'$ 
``suppression factor'' \cite{lep}.
First measurements of $\eta' \rightarrow \gamma \gamma^*$ decays 
have been performed at CLEO \cite{cleogamma}. The new WASA 4$\pi$ 
detector \cite{wasa} at CELSIUS will enable precision studies of 
$\eta$ and $\eta'$ decays.
Data expected in the next few years provides an exciting new opportunity
to study axial U(1) dynamics and to investigate the role of gluonic
degrees of freedom in $\eta$ and $\eta'$ physics.

In this lecture we focus primarily on $\eta'$ production in proton-proton
collisions together with a brief review of the axial U(1) problem in QCD.

The role of gluonic degrees of freedom and OZI violation in the 
$\eta'$--nucleon system has been investigated through the 
flavour-singlet Goldberger-Treiman relation \cite{venez,hatsuda},
the low-energy $pp \rightarrow pp \eta'$ reaction \cite{sb99} and
$\eta'$ photoproduction \cite{bww}.
The flavour-singlet Goldberger-Treiman relation connects 
the flavour-singlet 
axial-charge $g_A^{(0)}$ measured in polarised deep inelastic scattering 
with the $\eta'$--nucleon coupling constant $g_{\eta' NN}$.
Working in the chiral limit it reads
\begin{equation}
M g_A^{(0)} = \sqrt{3 \over 2} F_0 \biggl( g_{\eta' NN} - g_{QNN} \biggr) 
\end{equation}
where
$g_{\eta' NN}$ is the $\eta'$--nucleon coupling constant and $g_{QNN}$ 
is an OZI violating coupling which measures the one particle 
irreducible coupling of the topological charge density 
$Q = {\alpha_s \over 4 \pi} G {\tilde G}$ to the nucleon.
In Eq.(1) $M$ is the nucleon mass and $F_0$ ($\sim 0.1$GeV) renormalises 
\cite{sb99} the flavour-singlet decay constant.
The coupling constant $g_{QNN}$ is, in part, related \cite{venez} 
to the amount of spin carried by polarised gluons in a polarised proton.
The large mass of the $\eta'$ and the small value of $g_A^{(0)}$ 
\begin{equation}
\left. g^{(0)}_A \right|_{\rm pDIS} = 0.2 - 0.35
\end{equation}
extracted from deep inelastic scattering \cite{bass99,windmolders}
(about a $50\%$ OZI suppression) 
point to substantial violations of the OZI rule in the flavour-singlet 
$J^P=1^+$ channel \cite{okubo}.
A large positive $g_{QNN} \sim 2.45$
is one possible explanation of the small value of $g_A^{(0)}|_{\rm pDIS}$.

It is important to look for other observables which are sensitive to 
$g_{QNN}$.  
OZI violation in the $\eta'$--nucleon system is a probe of the role 
of gluons in dynamical chiral symmetry breaking in low-energy QCD.

Working with the $U_A(1)$--extended chiral Lagrangian for low-energy QCD 
\cite{vecca,lagran} 
--- see Section 3 below ---
one finds a gluon-induced contact interaction in the 
$pp \rightarrow pp \eta'$ reaction close to threshold \cite{sb99}:
\begin{equation}
{\cal L}_{\rm contact} =
         - {i \over F_0^2} \ g_{QNN} \ {\tilde m}_{\eta_0}^2 \
           {\cal C} \
           \eta_0 \ 
           \biggl( {\bar p} \gamma_5 p \biggr)  \  \biggl( {\bar p} p \biggr)
\end{equation}
Here ${\tilde m}_{\eta_0}$ is the gluonic contribution to the mass of the
singlet 0$^-$ boson and ${\cal C}$ is a second OZI violating coupling 
which also features in $\eta'N$ scattering.
The physical interpretation of the contact term (3) 
is a ``short distance'' ($\sim 0.2$fm) interaction 
where glue is excited in the interaction region of
the proton-proton collision and 
then evolves to become an $\eta'$ in the final state.
This gluonic contribution to the cross-section 
for $pp \rightarrow pp \eta'$ 
is extra to the contributions associated with 
meson exchange models \cite{holinde,wilkin,faldt,nakayama}.
There is no reason, a priori, to expect it to be small.

What is the phenomenology of this gluonic interaction ?

Since glue is flavour-blind the contact interaction (3) has the same 
size in both 
the $pp \rightarrow pp \eta'$ and $pn \rightarrow pn \eta'$ reactions.
CELSIUS \cite{celsius} have measured the ratio
$R_{\eta} 
 = \sigma (pn \rightarrow pn \eta ) / \sigma (pp \rightarrow pp \eta )$
for quasifree $\eta$ 
production from a deuteron target up to 100 MeV above threshold.
They observed that $R_{\eta}$ is approximately energy-independent 
$\simeq 6.5$ over the whole energy range --- see Fig.1.
\begin{figure}[ht]
\vspace{2.5truecm}
\epsfig{figure=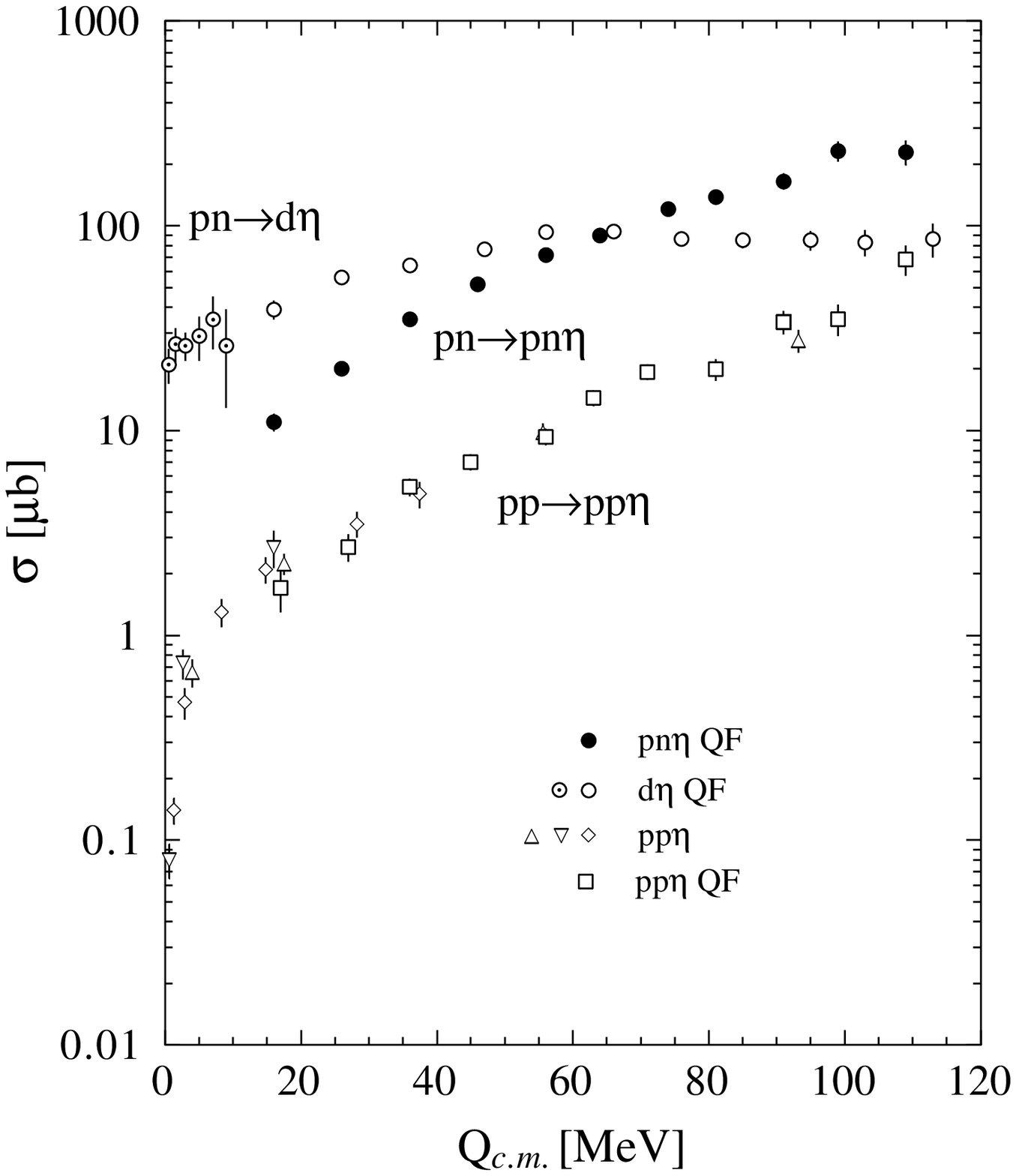, width=9.0cm, angle=0}
\caption{}
\label{stampc}
\end{figure}
%
The value of this ratio signifies a strong isovector exchange 
contribution to the $\eta$ production mechanism \cite{celsius}.
This experiment should be repeated for $\eta'$ production.
The cross-section for $pp \rightarrow pp \eta'$ 
close to threshold has been measured at COSY \cite{cosy}.
Following the suggestion in \cite{sb99}
a new COSY-11, Uppsala University Collaboration 
\cite{cosyprop}
has been initiated 
to carry out the $pn \rightarrow pn \eta'$ measurement.
The more important that the gluon-induced process (3) is in 
the $pp \rightarrow pp \eta'$ reaction the more one would expect 
$R_{\eta'} =
 \sigma (pn \rightarrow pn \eta' ) / \sigma (pp \rightarrow pp \eta' )$
to approach unity near threshold
after we correct for the final state 
interaction between the two outgoing nucleons.
(After we turn on the quark masses, the small $\eta-\eta'$ mixing angle 
 $\theta \simeq -18$ degrees means that the gluonic effect
 (3) should be considerably bigger in $\eta'$ production than $\eta$
 production.) 
$\eta'$ phenomenology is characterised by large OZI violations.
It is natural to expect 
large gluonic effects in the $pp \rightarrow pp \eta'$ process.

In Section 2 we give a brief Introduction to the U(1) problem. 
Section 3 introduces the chiral Lagrangian approach and 
Section 4 makes contact with the experimental data from CELSIUS and COSY.

\section{The U(1) problem}

In classical field theory Noether's theorem tells us that there is a 
conserved current associated with each global symmetry of the 
Lagrangian.
The QCD Lagrangian
\begin{equation}
{\cal L}_{QCD} = 
\sum_q
{\bar q}_L \biggl( i {\hat D} - g {\hat A} \biggr) q_L 
+
{\bar q}_R \biggl( i {\hat D} - g {\hat A} \biggr) q_R
- \sum_q m_q \biggl( {\bar q}_L q_R + {\bar q}_R q_L \biggr) 
- {1 \over 2} G_{\mu \nu} G^{\mu \nu}
\end{equation}
exhibits chiral symmetry for massless quarks: when the quark mass 
term is turned off the left- and right-handed quark fields do not
couple in the Lagrangian and transform independently under chiral 
rotations.

Chiral $SU(2)_L \otimes SU(2)_R$ 
\begin{equation}
\left(\begin{array}{c} 
u_L \\
d_L
\vphantom{\inv}  
\end{array}\right) 
\ \mapsto \
e^{i {1 \over 2}{\vec \alpha}.{\vec \tau} \gamma_5}
\left(\begin{array}{c} 
u_L \\
d_L 
\vphantom{\inv}  
\end{array}\right) 
\ \ \ \ \ , \ \ \ \ \ 
\left(\begin{array}{c} 
u_R \\
d_R
\vphantom{\inv}  
\end{array}\right) 
\ \mapsto \
e^{i {1 \over 2}{\vec \beta}.{\vec \tau} \gamma_5}
\left(\begin{array}{c} 
u_R \\
d_R 
\vphantom{\inv}  
\end{array}\right) 
\end{equation}
is associated with the isotriplet axial-vector current $J_{\mu 5}^{(3)}$
\begin{equation}
J_{\mu 5}^{(3)} =
\biggl[ {\bar u} \gamma_{\mu} \gamma_5 u 
      - {\bar d} \gamma_{\mu} \gamma_5 d
\biggr]
\end{equation}
which is partially conserved
\begin{equation}
\partial^{\mu} J_{\mu 5}^{(3)} =
2 m_u {\bar u} i \gamma_5 u - 2 m_d {\bar d} i \gamma_5 d
\end{equation}
The absence of parity doublets in the hadron spectrum tells us that
the near-chiral symmetry for light $u$ and $d$ quarks is spontaneously
broken.
Spontaneous chiral symmetry breaking is associated with a non-vanishing
chiral condensate
\begin{equation}
\langle \ {\rm vac} \ | \ {\bar q} q \ | \ {\rm vac} \ \rangle < 0
\end{equation}
The light-mass pion is identified as the corresponding Goldstone boson
and the current $J_{\mu 5}^{(3)}$ is associated with the pion through 
PCAC 
\begin{equation}
\langle {\rm vac} | J_{\mu 5}^{(3)}(z) | \pi (q) \rangle 
= -i f_{\pi} q_{\mu} e^{-i q.z}
\end{equation}
Taking the divergence equation
\begin{equation}
\langle {\rm vac} | \partial^{\mu} J_{\mu 5}^{(3)}(z) | \pi (q) \rangle 
= - f_{\pi} m_{\pi}^2 e^{-i q.z}
\end{equation}
the pion mass-squared vanishes in the chiral limit as $m_{\pi}^2 \sim m_q$.
This and PCAC \cite{dashen} are the starting points for chiral perturbation 
theory \cite{gasser}.

The non-vanishing chiral condensate also spontaneously 
breaks the axial U(1) symmetry so, naively, one might 
expect an isosinglet pseudoscalar degenerate with the pion. 
The lightest mass
isosinglet pseudoscalar is the $\eta$ meson which has a mass of 547 MeV.

The puzzle deepens when one considers SU(3).
Spontaneous chiral symmetry breaking suggests an octet of 
Goldstone bosons associated with chiral $SU(3)_L \otimes SU(3)_R$ 
plus a singlet boson associated 
with axial U(1) --- each with mass $m^2_{\rm Goldstone} \sim m_q$.
If the $\eta$ is associated with the octet boson then 
the Gell-Mann Okubo relation
\begin{equation}
m_{\eta_8}^2 = {4 \over 3} m_{\rm K}^2 - {1 \over 3} m_{\pi}^2
\end{equation}
is satisfied to within a few percent.
Extending the theory from $SU(3)$ to 
$SU(3)_L \otimes SU(3)_R \otimes U(1)$
the large strange quark mass induces considerable $\eta$-$\eta'$ mixing. 
Taking 
$m^2_{\rm Goldstone} \sim m_q$
the $\eta$ would be approximately an 
isosinglet light-quark state 
(${1 \over \sqrt{2}} | {\bar u} u + {\bar d} d \rangle$)
degenerate with the pion 
and the $\eta'$ would be approximately a strange quark state
$| {\bar s} s \rangle$ with mass about $\sqrt{2m_K^2 - m_{\pi}^2}$.
That is, the masses for the $\eta$ and $\eta'$ mesons 
with $\eta$-$\eta'$ mixing and without extra physical 
input come out about 300-400 MeV too small! This is the axial U(1) problem.

The extra physics which is needed to understand the U(1) problem 
are gluon topology and the QCD axial anomaly.
The (gauge-invariantly renormalised) 
flavour-singlet axial-vector current in QCD
satisfies the anomalous divergence equation 
\cite{adler,bell} 
\begin{equation}
\partial^\mu J_{\mu5} = 
\sum_{k=1}^{f} 2 i \biggl[ m_k \bar{q}_k \gamma_5 q_k 
\biggr]
+ N_f 
\biggl[ {\alpha_s \over 4 \pi} G_{\mu \nu} {\tilde G}^{\mu \nu} 
\biggr]
\end{equation}
where
\begin{equation}
J_{\mu 5} =
\left[ \bar{u}\gamma_\mu\gamma_5u
                  + \bar{d}\gamma_\mu\gamma_5d
                  + \bar{s}\gamma_\mu\gamma_5s \right]
\end{equation} 
Here $N_f=3$ is the number of light flavours, 
$G_{\mu \nu}$ is the gluon field tensor and
${\tilde G}^{\mu \nu} = {1 \over 2} \epsilon^{\mu \nu \alpha \beta}
G_{\alpha \beta}$.
The anomalous term
$Q(z) \equiv {\alpha_s \over 4 \pi} G_{\mu \nu} {\tilde G}^{\mu \nu} (z)$
is the topological charge density.
Its integral over space
$\int \ d^4 z \ Q = n$ 
measures the gluonic ``winding number'' \cite{rjc},
which is an integer for (anti-)instantons and which 
vanishes in perturbative QCD.
The exact dynamical mechanism how (non-perturbative) gluonic 
degrees of freedom contribute to axial U(1) symmetry breaking 
through 
the anomaly is still hotly debated 
\cite{christos,rjc,thooftrep,witteninst}:
suggestions include instantons 
\cite{thooft} and possible connections with confinement \cite{koguts}.

Independent of the detailed QCD dynamics one can construct low-energy 
effective chiral Lagrangians which include the effect of the anomaly 
and axial U(1) symmetry, 
and use these Lagrangians to study low-energy processes involving the 
$\eta$ and $\eta'$.

\section{The low-energy effective Lagrangian}

Starting in the meson sector, 
the building block for the $U_A(1)$-extended 
low-energy effective Lagrangian \cite{vecca,lagran} is
\begin{equation}
{\cal L}_{\rm m} = 
{F_{\pi}^2 \over 4} 
{\rm Tr}(\partial^{\mu}U \partial_{\mu}U^{\dagger}) 
+
{F_{\pi}^2 \over 4} {\rm Tr} \biggl[ \chi_0 \ ( U + U^{\dagger} ) \biggr]
+ 
{1 \over 2} i Q {\rm Tr} \biggl[ \log U - \log U^{\dagger} \biggr]
+ {3 \over {\tilde m}_{\eta_0}^2 F_{0}^2} Q^2 .
\end{equation}
Here
\begin{equation}
U = \exp \ \biggl(  i {\phi \over F_{\pi}}  
                  + i \sqrt{2 \over 3} {\eta_0 \over F_0} \biggr) 
\end{equation}
is the unitary meson matrix where $\phi = \sum_k \phi_k \lambda_k$
with $\phi_k$ 
denotes the octet of would-be Goldstone bosons 
$(\pi, K, \eta_8)$
associated with 
spontaneous chiral 
$SU(3)_L \otimes SU(3)_R$ breaking, $\eta_0$ 
is the singlet boson and $Q$ is the topological charge density;
$\chi_0 = 
{\rm diag} [ m_{\pi}^2, m_{\pi}^2, (2 m_K^2 - m_{\pi}^2 ) ]$
is the meson mass matrix.
The pion decay constant $F_{\pi} = 92.4$MeV and 
$F_0$ renormalises the flavour-singlet decay constant.

When we expand out the Lagrangian (14) the first term 
contains the kinetic energy term for the pseudoscalar 
mesons;
the second term contains the meson mass terms before 
coupling to gluonic degrees of freedom.
The $U_A(1)$ gluonic potential involving the topological 
charge density is constructed to reproduce 
the axial anomaly (12) in the divergence of 
the gauge-invariantly renormalised axial-vector current
and to generate the gluonic contribution to the $\eta$ 
and $\eta'$ masses.
The gluonic term $Q$ is treated as a background field with no kinetic 
term.
It may be eliminated through its equation of motion
\begin{equation}
{1 \over 2} i Q {\rm Tr} \biggl[ \log U - \log U^{\dagger} \biggr]
+ {3 \over {\tilde m}_{\eta_0}^2 F_{0}^2} Q^2 
\
\mapsto \
- {1 \over 2} {\tilde m}_{\eta_0}^2 \eta_0^2 
\end{equation}
making the gluonic mass term clear.
After $Q$ is eliminated from the effective Lagrangian 
via (16),
we expand ${\cal L}_{\rm m}$ to ${\cal O}(p^2)$ 
in momentum keeping finite quark masses and obtain:
\begin{eqnarray}
{\cal L}_{\rm m} = & &
\sum_k {1 \over 2} \partial^{\mu} \phi_k \partial_{\mu} \phi_k 
+
{1 \over 2} \partial_{\mu} \eta_0 \partial^{\mu} \eta_0 
\ \biggl( {F_{\pi} \over F_0} \biggr)^2 \ 
- {1 \over 2} {\tilde m}_{\eta_0}^2 \eta_0^2 \ 
\\ \nonumber
&-& {1 \over 2} 
m_{\pi}^2 \biggl( 2 \pi^+ \pi^- + \pi_0^2 \biggr)
- m_K^2 \biggl( K^+ K^- + K^0 {\bar K}^0 \biggr)
- {1 \over 2} \biggl( {4 \over 3} m_K^2 - {1\over 3} m_{\pi}^2 \biggr)
  \eta_8^2
\\ \nonumber
&-& {1 \over 2} \biggl( {2 \over 3} m_K^2 + {1\over 3} m_{\pi}^2 \biggr)
    \ \biggl( {F_{\pi} \over F_0} \biggr)^2 \ \eta_0^2
    + {4 \over 3 \sqrt{2}} \biggl( m_K^2 - m_{\pi}^2 \biggr) 
    \ \biggl( {F_{\pi} \over F_0} \biggr) \eta_8 \eta_0 + ...
\end{eqnarray}

The value of $F_0$ is usually determined from the decay rate for
$\eta' \rightarrow 2 \gamma$.
In QCD 
one finds the relation \cite{shore}
\begin{equation}
{2 \alpha \over \pi} = 
\sqrt{3 \over 2} F_0 \biggl( g_{\eta' \gamma \gamma} - g_{Q \gamma \gamma} 
\biggr) 
\end{equation}
(in the chiral limit)
which is derived by coupling the effective Lagrangian (14) to photons.
The observed decay rate \cite{twogamma} is consistent \cite{gilman} 
with the OZI prediction 
for $g_{\eta' \gamma \gamma}$ 
{\it if} $F_0$ and $g_{Q \gamma \gamma}$ take their 
OZI values: 
$F_0 \simeq F_{\pi}$ and $g_{Q \gamma \gamma} = 0$.
Motivated by this observation it is common to take $F_0 \simeq F_{\pi}$.

\subsection{Glue and the $\eta$ and $\eta'$ masses}

If we work in the approximation $m_u = m_d$ and set $F_0 = F_{\pi}$, 
then the $\eta - \eta'$ mass matrix which follows from (17) becomes
\begin{equation}
M^2_{\eta - \eta'} =\
\left(\begin{array}{cc} 
{4 \over 3} m_{\rm K}^2 - {1 \over 3} m_{\pi}^2  &
- {2 \over 3} \sqrt{2} (m_{\rm K}^2 - m_{\pi}^2) \\
\\
- {2 \over 3} \sqrt{2} (m_{\rm K}^2 - m_{\pi}^2) &
[ {2 \over 3} m_{\rm K}^2 + {1 \over 3} m_{\pi}^2 + {\tilde m}^2_{\eta_0} ]
\vphantom{\inv}  
\end{array}\right) 
\end{equation}
with $\eta$-$\eta'$ mixing 
\begin{eqnarray}
| \eta \rangle &=& 
\cos \theta \ | \eta_8 \rangle - \sin \theta \ | \eta_0 \rangle
\\ \nonumber
| \eta' \rangle &=& 
\sin \theta \ | \eta_8 \rangle + \cos \theta \ | \eta_0 \rangle
\end{eqnarray}
driven predominantly by the large strange-quark mass.
The Gell-Mann Okubo mass formula (11) can be seen in the top left 
matrix element of the mass matrix (19).
Diagonalising the $\eta$--$\eta'$ mass matrix we obtain values 
for the $\eta$ and $\eta'$ masses:
\begin{equation}
m^2_{\eta', \eta} = (m_{\rm K}^2 + {\tilde m}_{\eta_0}^2 /2) 
\pm {1 \over 2} 
\sqrt{(2 m_{\rm K}^2 - 2 m_{\pi}^2 - {1 \over 3} {\tilde m}_{\eta_0}^2)^2 
   + {8 \over 9} {\tilde m}_{\eta_0}^4} .
\end{equation}
If we turn off the gluon mixing term, 
then one finds
$m_{\eta'} = \sqrt{2 m_{\rm K}^2 - m_{\pi}^2}$ 
and
$m_{\eta} = m_{\pi}$.
Without any extra input from glue, in the OZI limit, 
the $\eta$ would be approximately an isosinglet light-quark state 
(${1 \over \sqrt{2}} | {\bar u} u + {\bar d} d \rangle$)
degenerate with the pion and 
the $\eta'$ would a strange-quark state $| {\bar s} s \rangle$
--- mirroring the isoscalar vector $\omega$ and $\phi$ mesons.
Indeed, in an early paper \cite{weinberg} Weinberg argued that
the mass of the $\eta$ would be less than $\sqrt{3} m_{\pi}$
without any extra U(1) dynamics to further break the axial U(1)
symmetry.
Summing over the two eigenvalues in (21) yields \cite{vecca} 
\begin{equation}
m_{\eta}^2 + m_{\eta'}^2 = 2 m_K^2 + {\tilde m}_{\eta_0}^2 .
\end{equation}
Substituting the physical values of $(m_{\eta}^2 + m_{\eta'}^2)$ 
in 
Eq.(22) and $m_K^2$ yields ${\tilde m}_{\eta_0}^2 = 0.73$GeV$^2$,
which corresponds to
$m_{\eta} = 499$MeV and $m_{\eta'} = 984$MeV.
The value ${\tilde m}_{\eta_0}^2 = 0.73$GeV$^2$ 
corresponds to an $\eta - \eta'$ 
mixing angle $\theta \simeq - 18$ degrees
--- 
which is within the range -17 to -20 degrees obtained 
from a
study of various decay processes in \cite{gilman,frere}.
The physical masses are $m_{\eta} = 547$MeV and $m_{\eta'} = 958$MeV.
Closer agreement with the physical masses can be obtained by taking
$F_0 \neq F_{\pi}$ and including higher-order mass terms in the chiral
expansion.
Two mixing angles \cite{leutwyler,feldmann} enter the $\eta - \eta'$ 
system when one extends the theory and ${\cal L}_{\rm m}$ to $O(p^4)$ 
in the meson momentum.
(The two mixing angles are induced by $F_{\pi} \neq F_K$ due to chiral
 corrections at $O(p^4)$ \cite{gasser}.)

\subsection{OZI violation and the $\eta'$--nucleon interaction}

The low-energy effective Lagrangian (14) is readily extended to include 
$\eta$--nucleon and $\eta'$--nucleon coupling.
Working to $O(p)$ in the meson momentum the chiral Lagrangian 
for meson-baryon coupling is 
\begin{eqnarray}
{\cal L}_{\rm mB} &=&
{\rm Tr} \ {\overline B} (i \gamma_{\mu} D^{\mu} - M_0) B
\\ \nonumber
&+& F \
{\rm Tr} \biggl( {\overline B} \gamma_{\mu} \gamma_5 [a^{\mu}, B]_{-}
  \biggr)
+ D \ 
{\rm Tr} \biggl( {\overline B} \gamma_{\mu} \gamma_5 \{a^{\mu}, B\}_{+}
  \biggr)
\\ \nonumber
&+&
{i \over 3}
K \ {\rm Tr} \biggl({\overline B} \gamma_{\mu} \gamma_5 B \biggr)
    {\rm Tr} \biggl(U^{\dagger} \partial^{\mu} U \biggr) 
- {{\cal G}_{QNN} \over 2 M_0} \partial^{\mu} Q 
  {\rm Tr} \biggl( {\overline B} \gamma_{\mu} \gamma_5 B \biggr) 
+ 
{{\cal C} \over F_0^4} Q^2 {\rm Tr} \biggl( {\overline B} B \biggr) 
\end{eqnarray}
Here
\begin{equation}
B =\
\left(\begin{array}{ccc} 
{1 \over \sqrt{2}} \Sigma^0 + {1 \over \sqrt{6}} \Lambda & \Sigma^+ & p \\
\\
\Sigma^- & -{1 \over \sqrt{2}} \Sigma^0 + {1\over \sqrt{6}} \Lambda & n \\
\\
\Xi^- & \Xi^0 & -{2 \over \sqrt{6}} \Lambda
\vphantom{\inv}  
\end{array}\right) 
\end{equation}
denotes the baryon octet and $M_0$ denotes the baryon mass in the chiral 
limit.
In Eq.(23) $D_{\mu}$ is the chiral covariant derivative and
$
a_{\mu} = 
- {1 \over 2 F_{\pi}} \partial_{\mu} \phi 
- {1 \over 2 F_0} \sqrt{2 \over 3} \partial_{\mu} \eta_0 + ...
$
is the axial-vector current operator.
The SU(3) couplings are
$F= 0.459 \pm 0.008$ and $D= 0.798 \pm 0.008$ \cite{fec}.
The Pauli-principle forbids any flavour-singlet $J^P={1 \over 2}^+$ 
ground-state baryon degenerate with the baryon octet $B$.
In general, one may expect OZI violation wherever a coupling involving
the $Q$-field occurs.

Following Eq.(16), we eliminate $Q$ from the total Lagrangian 
${\cal L} = {\cal L}_{\rm m} + {\cal L}_{\rm mB}$
through its equation of motion.
The $Q$ dependent terms in the effective Lagrangian become:
\begin{eqnarray}
{\cal L}_Q &=& {1 \over 12} {\tilde m}_{\eta_0}^2 \
   \biggl[ \ - 6 \eta_0^2 \  
           - \ {\sqrt{6} \over M_0} \ {\cal G}_{QNN} \ F_0 \ 
               \partial^{\mu} \eta_0 \
           {\rm Tr} \biggl( {\bar B} \gamma_{\mu} \gamma_5 B \biggr) 
\\ \nonumber
& & \ \ \ \ \ \ \ \ 
           + \ {\cal G}_{QNN}^2 \ F_0^2 \
             \biggl( {\rm Tr} {\bar B} \gamma_5 B \biggr)^2 \
           + \ 2 \ {\cal C} \ {{\tilde m}_{\eta_0}^2 \over F_0^2 } \
           \eta_0^2 \ {\rm Tr} \biggl( {\bar B} B \biggr) 
\\ \nonumber
& & \ \ \ \ \ \ \ \ 
           - \ {\sqrt{6} \over 3 M_0 F_0} \ {\cal G}_{QNN} \
           {\cal C} {\tilde m}_{\eta_0}^2 
           \eta_0 \ \partial^{\mu}  
           {\rm Tr} \biggl( {\bar B} \gamma_{\mu} \gamma_5 B \biggr)  \
           {\rm Tr} \biggl( {\bar B} B \biggr)
+ ... \biggr] 
\end{eqnarray}
This equation describes the gluonic contributions to 
the $\eta$-nucleon and $\eta'$-nucleon interactions.
The term
$ - \ {\sqrt{6} \over M_0} \ {\cal G}_{QNN} \ F_0 \ 
               \partial^{\mu} \eta_0 \
           {\rm Tr} \biggl( {\bar B} \gamma_{\mu} \gamma_5 B \biggr)$
is a gluonic (OZI violating) contribution 
to the $\eta'$--nucleon coupling constant, 
which is 
$
g_{\eta_0 NN}
= \sqrt{2 \over 3} {m \over F_0} 
  ( 2D + 2 K + {\cal G}_{QNN} F_0^2 { {\tilde m}^2_{\eta_0} \over 2 m } ) 
$
in the notation of (23).
The Lagrangian (25) has three contact terms associated with the gluonic 
potential in $Q$.
We recognise
$
{\cal L}^{(2)}_{\rm contact} =
         - {\sqrt{6} \over 12 m F_0} {\cal G}_{QNN} {\tilde m}_{\eta_0}^2 \
           {1 \over 3} {\cal C} {\tilde m}_{\eta_0}^2 \
           \eta_0 \ \partial^{\mu} 
           {\rm Tr} \biggl( {\bar B} \gamma_{\mu} \gamma_5 B \biggr)  \
           {\rm Tr} \biggl( {\bar B} B \biggr)
$
as the 
gluonic contact term (3) in the low-energy $pp \rightarrow pp \eta'$ 
reaction
with
$g_{QNN} \equiv
 \sqrt{1 \over 6} {\cal G}_{QNN} F_0 { \tilde m}^2_{\eta_0}$.
The term
$
{\cal L}^{(3)}_{\rm contact} =
  {1 \over 6 F_0^2} \ {\cal C} \ 
  {\tilde m}_{\eta_0}^4 \ \eta_0^2 \ {\rm Tr} \biggl( {\bar B} B \biggr)
$
is potentially important to $\eta$--nucleon and $\eta'$--nucleon scattering 
processes. 
The contact terms
${\cal L}^{(j)}_{\rm contact}$ are proportional to ${\tilde m}_{\eta_0}^2$
($j=2$) and ${\tilde m}_{\eta_0}^4$ ($j=3$) which vanish in the formal OZI 
limit.
Phenomenologically, the large masses of the $\eta$ and $\eta'$ 
mesons means 
that there is no reason, a priori, to expect the 
${\cal L}^{(j)}_{\rm contact}$ to be small.

Gluonic $U_A(1)$ degrees of freedom induce several ``$\eta'$--nucleon
coupling constants''.
The three couplings 
($g_{\eta_0 NN}$, ${\cal G}_{QNN}$ and ${\cal C}$) are each 
potentially important in the theoretical description the 
$\eta'$--nucleon and $\eta'$--two-nucleon systems.
Different combinations of these coupling constants are relevant to 
different $\eta'$ production processes and to the flavour-singlet 
Goldberger-Treiman relation.
Testing the sensitivity of $\eta'$--nucleon interactions to the gluonic 
terms in the effective chiral Lagrangian for low-energy QCD will teach 
us about the role of gluons in chiral dynamics.

\section{Proton-proton collisions}

How important is the contact interaction ${\cal L}^{(2)}_{\rm contact}$ 
in the $pp \rightarrow pp \eta'$ reaction ?

The T-matrix for $\eta'$ production in proton-proton collisions,
$p_1 (\vec{p}) + p_2 (-\vec{p}) \rightarrow p + p + \eta'$, 
at threshold in the centre of mass frame is
\begin{equation}
{\rm T^{cm}_{th}} (pp \rightarrow pp \eta') 
=
{\cal A} 
\biggl[ i ( \vec{\sigma}_1 - \vec{\sigma}_2 )
        + \vec{\sigma}_1 {\rm x} \vec{\sigma}_2 \biggr].{\vec p}
\end{equation}
where
${\cal A}$ is the (complex) threshold amplitude for $\eta'$ production.
Measurements of the total cross-section for $pp \rightarrow pp \eta'$
have been published by COSY \cite{cosy} and 
SATURNE \cite{saturne} 
between 1.5 and 24 MeV above threshold -- see Fig.2.
\begin{figure}[h]
\vspace{1.5truecm}
\epsfig{figure=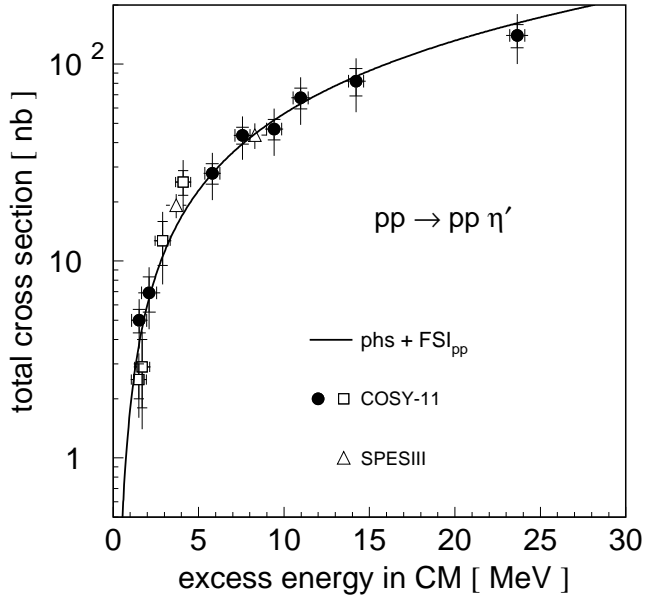, width=10.0cm, angle=0}
\caption{The COSY and SATURNE data on $pp \rightarrow pp \eta'$}
\label{stampb}
\end{figure}


The energy dependence of the data are well described by phase space plus
proton-proton final state interaction (neglecting any $\eta'$-p FSI).
Using the model of Bernard et al.
\cite{norbert}
treating the $pp$ final state interaction in effective range approximation
one finds a good fit to the measured total cross-section data with
\begin{equation}
|{\cal A}| = 0.21 \ {\rm fm}^4.
\end{equation}

The present (total cross-section only) data on $pp \rightarrow pp \eta'$
is insufficient to distinguish between possible production 
mechanisms involving the (short-range) gluonic contact term (3) 
and the 
long-range contributions associated with meson exchange models.
Long-range meson exchange contributions to ${\cal A}$ involve 
the exchange of a $\pi^0$, $\eta$, $\omega$ or $\rho^0$ 
between the two protons and the emission of an $\eta'$ from
one of the two protons. 
This process involves $g_{\eta_0 NN}$.
The contact term (3) involves the excitation of gluonic degrees of 
freedom in the interaction region, 
is isotropic and involves the product of 
${\cal G}_{QNN}$ and the second gluonic coupling ${\cal C}$.
In their analysis of the SATURNE data on $pp \rightarrow pp \eta'$
Hibou et al. \cite{saturne} found that a one-pion exchange model 
adjusted to fit the S-wave contribution 
to the $pp \rightarrow pp \eta$ cross-section near threshold yields 
predictions about 30\% below the measured $pp \rightarrow pp \eta'$ 
total cross-section.
The gluonic contact term (3) 
is a candidate for additional, potentially important, short range interaction.

To estimate how strong the contact term must be in order to make an 
important contribution to the measured $pp \rightarrow pp \eta'$ 
cross-section, let us consider the extreme scenario where the value 
of $|{\cal A}|$ in Eq.(27) is saturated by the contact term (3).
If we take the estimate $g_{QNN} \sim 2.45$
(or equivalently ${\cal G}_{QNN} \sim +60$GeV$^{-3}$)
suggested by the polarised deep inelastic scattering and 
the flavour-singlet Goldberger-Treiman relation below Eq.(2),
then we need
${\cal C} \sim 1.8$GeV$^{-3}$
to saturate $|{\cal A}|$.
The OZI violating parameter 
${\cal C} \sim 1.8$GeV$^{-3}$ seems 
reasonable compared with ${\cal G}_{QNN} \sim 60$GeV$^{-3}$.

To help resolve the different production mechanisms it will be important 
to test the isospin dependence of the 
$pN \rightarrow pN \eta'$ process
through quasi-free production from the deuteron \cite{sb99,cosyprop} and 
to make a partial wave analysis of the $\eta'$ production process,
following the work pioneered by CELSIUS for $\eta$ production \cite{partial}.
Here,
it is interesting to note that the recent higher-energy 
($p_{\rm beam} = 3.7$GeV)
measurement of the $pp \rightarrow pp \eta'$ cross-section by 
the DISTO collaboration
\cite{disto} suggests isotropic $\eta'$ production at this energy.

\clearpage

{\bf Acknowledgements} \\

It is a pleasure to thank P. Moskal for organising this stimulating 
workshop in the beautiful surroundings of Cracow.
I thank E. Gabathuler, T. Johansson, S. Kullander, P. Moskal, G. Rudolph,
J. Stepaniak and U. Wiedner for helpful communications about experimental 
data.

\vspace{1.0cm}


\end{document}